\documentstyle[12pt]{article}
\title{\bf A quantum cosmology and discontinuous signature changing classical solutions }
\author{F. Darabi \thanks{e-mail:f.darabi@azaruniv.edu} and
A. Rastkar \thanks{e-mail:a$_-$rastkar@azaruniv.edu}\\
{\small Department of Physics, Azarbaijan University of Tarbiat
Moallem, 53714-161 Tabriz, Iran .} }
\begin{document}
\maketitle
\begin{abstract}
We revisit the classical and quantum cosmology of a universe in
which a self interacting scalar field is coupled to gravity with a
flat FRW type metric undergoing continuous signature transition.
We arrange for quantum cosmologically allowed discontinuity in the
classical solutions at the signature changing hypersurface,
provided these solutions be dual in some respects. This may be of
some importance in the study of early universe within the
signature changing scenarios.

\end{abstract}

\newpage

\section{Introduction}

The initial idea of signature change is due to Hartle and Hawking
\cite{HHS}. This idea makes it possible to have both Euclidean and
Lorentzian metrics in the path integral approach to quantum
gravity. It was later shown that signature change may happen even
in classical general relativity \cite{CSC}-\cite{Kos}. More
recently, the people have studied this issue in the Brane-World
scenario, as well \cite{BW}. From a classical point of view, the
signature change may prevent the occurrence of singularities in
general relativity, such as the Big Bang, which may be replaced by
a compact Euclidean domain prior to the birth of time in the
Lorentzian domain, the so-called no-boundary proposal \cite{HHS}.
Alternatively, the classical signature change scenario may be an
effective classical description of the quantum tunnelling approach
for the creation of Lorentzian universe \cite{QT}.

In general, there are two different approaches to the issue of
classical signature change: continuous and discontinuous. In the
continuous approach, passing from Euclidean to Lorentzian domain,
the signature of metric changes continuously, hence the metric
becomes degenerate at the transition hypersurface. In the
discontinuous approach however, the metric is non-degenerate
everywhere and discontinuous at the transition hypersurface. In
both approaches the dynamical fields and their first derivatives
satisfy specific junction conditions. In Ellis {\it et al} point
of view \cite{CSC}, both the fields and their first derivatives
are continuous, while in that of Hayward \cite{Hay} the fields are
continuous but their derivatives are zero at the transition
hypersurface.

In this paper, we first review the model adopted by Dereli and
Tucker \cite{DT} in which a self interacting scalar field is
coupled to gravity. In the classical version of this model,
Einstein field equations are solved such that the scalar field and
the scale factor are considered as dynamical variables, giving
rise to cosmological solutions with degenerate metrics describing
continuous transition from Euclidean to Lorentzian domains. The
quantum cosmological version of this model is also used to derive
the wavefunction of the universe by solving the corresponding
Wheeler-DeWitt equation \cite{DOT}. The Wheeler-DeWitt equation is
obtained by adopting a new choice of variables, through which
Einstein classical equations of motion arise from an anisotropic
constrained oscillator-ghost-oscillator Hamiltonian. A family of
Hilbert subspaces are then derived in which states are identified
with non dispersive wave packets which remarkably peak in the
vicinity of classical loci with parameterizations corresponding to
metric solutions of Einstein equations that admit a continuous
signature transition \cite{DOT}.

Then, we pay more attention to the previously obtained duality
transformations, on the parameters of the scalar field potential,
by which the classical cosmology transforms to its {\it dual}, but
the quantum cosmology remains unchanged \cite{FD}. Since these
duality transformations do not affect the results in \cite{DOT},
then a remarkable correspondence can exist between the self dual
quantum cosmology and the dual classical cosmologies. This
motivates the definition of a potential distribution over the
entire manifold such that Euclidean and Lorentzian parts are
endowed by dual potentials. Therefore, according to the above
correspondence, the dual classical cosmologies are
indistinguishable at the quantum level, and the jump from one
solution in the Euclidean part to its dual in the Lorentzian part
is consistent with this quantum cosmology.

Therefore, in spite of the common continuity requirement on the
signature changing classical solutions \cite{CSC}-\cite{Kos}, the
quantum cosmology discussed here allows for discontinuity in the
corresponding classical solutions passing through the hypersurface
of signature change. Of course, this is not so surprising because
the quantum mechanics always changes a continuous picture to a
discontinuous one.

We shall discuss on the possible interpretation of these allowed
jumps (supported by quantum cosmology) as sudden birth of a
Lorentzian universe from a rather small size Euclidean region. It
is to be noted that this interpretation is in no way equal to an
inflationary model. But it deserves to be studied to see how the
sudden creation of a Lorentzian universe may be accommodated in a
simple quantum cosmology admitting classical signature transition.

\section{Dual classical solutions}

We consider the Einstein-Hilbert action
\begin{equation}
I=\int_{}^{}\!\sqrt{|g|}\: \left[\frac{1}{16 \pi G} {\cal R} +
{\cal L}_M\right] d^{4}x,
\end{equation}
where ${\cal R}$ is the scalar curvature, ${\cal L}_M=\frac{1}{2}
\partial_{0}\phi\:
\partial^{0}\phi - U(\phi)$ is the real scalar field Lagrangian and $\phi$ is assumed to be
a homogeneous field which depends merely on the time parameter. We
take the chart $\{\beta,x^1,x^2,x^3\}$ and parameterize the metric
as FRW type \cite{DT}
\begin{eqnarray}
g = -\beta d\beta \otimes d\beta + \frac{R^2
(\beta)}{[1+{\frac{k}{4} r^2}]^2} \sum_i d{x^i} \otimes d{x^i},
\label{eq3}
\end{eqnarray}
where $r^2=\sum x_ix^i$, $R(\beta )$ is the scale factor with
$k=\{-1,0,1\} $ representing open, flat or closed universes and
$\beta $ is the lapse function producing the hypersurface of
signature change at $\beta =0$. For $\beta
>0$, the cosmic time can be written as $t=\frac 23\beta ^{3/2}$. By calculating the scalar
curvature ${\cal R}$ and using the transformations \cite{DT}
\begin{eqnarray}
X &=& R^{3/2} \cosh(\alpha \phi),  \label{eq8}\\
Y &=& R^{3/2} \sinh(\alpha \phi),  \label{eq9}
\end{eqnarray}
the corresponding effective Lagrangian is obtained for a flat
universe $k=0$\footnote{The models with $k\neq 0$ was also studied
in \cite{GGS}.}
\begin{equation}
{\cal L} = \frac{1}{2{\alpha}^{2}}(-\dot{X}^2 + \dot{Y}^2) - (X^2
- Y^2) U(\phi(X,Y)).
\label{eq10'}
\end{equation}
We know from the (3+1) decomposition of the Einstein theory of
gravity, that the Hamiltonian ${\cal H}$ corresponding to ${\cal
L}$, must vanish identically leading to a {\it zero energy
condition}
\begin{equation}
{\cal H} = \frac{1}{2{\alpha}^{2}}(-\dot{X}^2 + \dot{Y}^2) + (X^2
- Y^2) U(\phi(X,Y))=0. \label{eq10''}
\end{equation}
Now, we take the potential as \cite{DT}, \cite{FD}
\begin{equation}
\label{eq14}U(\phi )=\lambda +\frac 1{2{\alpha }^2}m^2\sinh
^2(\alpha \phi )+\frac 1{2{\alpha }^2}b\sinh (2\alpha \phi ),
\label{10}
\end{equation}
where $\lambda =U\mid _{\phi =0}$, $m^2=\frac{{%
\partial }^2U}{\partial {\phi }^2}\mid _{\phi =0}$ and $b$
are the bare cosmological constant, positive mass square and
coupling constant, respectively. The minimum of the potential
occurs at
$\phi=-\frac{1}{2\alpha}\tanh^{-1}(\frac{2b}{m^2})$ when $|{2b/m^2}%
|<1 $ and we will interpret this minimum as the effective
cosmological constant
\begin{equation}
\Lambda_{eff} = \lambda+\frac{m^2}{4\alpha^2}\left(\sqrt{1-\frac{4
b^2}{m^4}}-1\right)_, \label{10'}
\end{equation}
where the second term is interpreted as the contribution of the
field excitation. Variation of the action with respect to the
dynamical variables X and Y gives the dynamical equations
\begin{equation}
\ddot{\xi}=M \xi,
\label{9}
\end{equation}
subjected to the zero energy condition
\begin{equation}
\dot{\xi}^T J \dot{\xi}=\xi^T JM \xi, \label{11}
\end{equation}
where $\xi=\left(\begin{array}{c} X \\ Y
\end{array}\right), J=\left(\begin{array}{cc} 1 & 0
\\ 0  & -1 \end{array}\right)$, and
$$
M=\left(\begin{array}{cc} 2\alpha^2\lambda & b
\\ -b  & 2\alpha^2\lambda -m^2
\end{array}\right)_.
$$
One may then define the normal modes $\alpha=S^{-1} \xi$ by
\begin{equation}
\alpha=\left(\begin{array}{c} u \\ v
\end{array}\right) \:\: , \:\:   S=\left(\begin{array}{cc} -\frac{m^2-\sqrt{m^4-4b^2}}{2b} &
-\frac{m^2+\sqrt{m^4-4b^2}}{2b}
\\ \\ 1  & 1
\end{array}\right)_,
\label{12}
\end{equation}
which diagonalize the matrix $M$ as follows
$$
S^{-1}MS=\left(\begin{array}{cc} \lambda_+ & 0
\\ 0  & \lambda_- \end{array}\right)_.
$$
The zero energy condition then reads as
\begin{equation}
\dot{\alpha}^T {\cal J} \dot{\alpha} = \alpha^T {\cal I} \alpha,
\label{13}
\end{equation}
where ${\cal J}=S^T J S$ and ${\cal I}=S^{T}JMS$. The normal modes
$\left(\begin{array}{c}
u \\
v
\end{array}
\right)$, with zero energy condition (\ref{13}) evaluated at
$t=0$, for $\lambda _{+},\lambda _{-}<0$ lead to the following
classical loci \cite{DOT}
\begin{eqnarray}
v&=& 2 A_0 \cos\left[ \frac{1}{r}
\mbox{cos}^{-1}\left(\epsilon\frac{ru}{2A_0}\right)\right],
\hspace{4mm}
|u|\leq\frac{2A_0}{r}, \nonumber\\
   \label{eq15}\\
v&=& 2 A_0 \cosh\left[ \frac{1}{r}
\mbox{cosh}^{-1}\left(\epsilon\frac{ru}{2A_0}\right)\right],
\hspace{4mm} |u|>\frac{2A_0}{r}, \nonumber
\end{eqnarray}
where $r=\sqrt{\frac{\lambda _{+}}{\lambda _{-}}}$ , $0<r<1$,
 $\epsilon =\pm 1$ indicates
two ways of satisfying the constraint ${\cal H}=0$, and $\lambda
_{\pm }$ are the eigenvalues of the decoupling matrix given by
\begin{equation}
\lambda _{\pm }=\frac{3\lambda }4-\frac{m^2}2\pm \frac 12\sqrt{%
m^4-4b^2}.\label{eq18'}
\end{equation}
An interesting feature of this model is that one can find a class
of transformations on the space of parameters $\{\lambda, m^2,
b\}$ leaving the eigenvalues $\lambda _{\pm }$ of the decoupling
matrix invariant \cite{FD}. These transformations can be written
as
\begin{eqnarray}
& &\lambda \rightarrow \tilde{\lambda} \equiv \frac{1}{4 \alpha^4} \lambda^{-1}, \nonumber\\
& &m^2 \rightarrow \tilde{m}^2 \equiv m^2 - \frac{4 \alpha^4 \lambda^2 -1}{\alpha^2 \lambda}, \label{eq18}\\
& &b^2 \rightarrow \tilde{b}^2 \equiv b^2 + m^2 [(2 \alpha^2
\lambda)^{-1} - 2 \alpha^2 \lambda] + [(2 \alpha^2 \lambda)^{-1} -
2 \alpha^2 \lambda]^2. \nonumber
\end{eqnarray}
It is seen that although the classical loci (\ref{eq15}) do not
change under (\ref{eq18}), the corresponding solutions $R(\beta )$
and $\phi (\beta )$
change, since $X(\beta )$ and $Y(\beta )$ are related to $u(\beta )$ and $%
v(\beta )$ by the decoupling matrix which is changed under
(\ref{eq18}).

Therefore, if we define (\ref{eq18}) as {\it duality}
transformations, then we have two sets of solutions for $R(\beta
)$ and $\phi (\beta )$ corresponding to dual sets of physical
parameters. We interpret the new parameters as dual bare
cosmological constant, dual mass square and dual coupling
constant, respectively.

As is discussed in \cite{DT}, in order to have signature
transition from Euclidean to Lorentzian, both eigenvalues $\lambda
_{\pm }$ must be negative, so equation (\ref{eq18'}) gives
\begin{equation}
\lambda <\frac 43[\frac{m^2}2-\frac 12\sqrt{m^4-4b^2}],
\end{equation}
but it does not guarantee that the dual potential has also a
minimum and a positive mass square. In order the dual potential
has these features, as well, we take
\begin{equation}
\label{eq19}\frac{m^2}2-\frac 12\sqrt{m^4-4b^2}\leq 1.
\end{equation}
If we now choose the set of suitably small couplings $\{\lambda,
m^2, b\}$
\begin{equation}
\lambda \simeq 0 \hspace{20mm}m^2 \ll 1 \hspace{20mm} b \simeq 0,
\end{equation}
with $b \ll m^2$, then $\mid 2b/m^2\mid <1$ and the potential $U$
will have a minimum. The dual transformations map the small values
of $\lambda, m^2 $ and $ b $ to very large values of the
corresponding dual parameters, $\tilde \lambda, \tilde{m}^2 $ and
$\tilde b$. Fortunately, the large values for $\tilde b$ and
$\tilde{m}^2 $ satisfy $\mid 2\tilde b/{\tilde m}^2\mid <1$ so
that we have a minimum for $\tilde U$, as well \cite{FD}. It then
follows that two different classical cosmologies, one with very
small values for the bare cosmological constant, mass scale and
coupling constant and the other with large ones, exhibit the same
signature dynamics on the configuration space $(u,v)$.

\section{A self-dual quantum cosmology}

We have shown, in this model, that it is possible to find dual
classical cosmologies ($R, \phi $)  and $(\tilde{R},
\tilde{\phi})$ corresponding to the same classical cosmology
defined on the ($u, v$) configuration space. On the other hand, it
was shown that the Wheeler-Dewitt equation for this model in the
mini-superspace ($u, v$) is \cite{DOT}
\begin{equation}
\left\{ \frac{\partial ^2}{\partial u^2}-\frac{\partial
^2}{\partial v^2}-\omega _1^2u^2+\omega _2^2v^2\right\} \Psi
(u,v)=0,
\end{equation}
where $\omega _1^2=-\lambda _{+},\omega _2^2=-\lambda _{-}$. To
obtain the solutions $\Psi^{(m_1, m_2)}(u,v)$ a quantization
condition was required
\begin{equation}
\frac{\lambda _{+}}{\lambda
_{-}}=\left(\frac{2m_1+1}{2m_2+1}\right)^2, \label{123}
\end{equation}
which was imposed on the parameters of the scalar field potential.
For a given pair of ($m_1,m_2$) it was shown by graphical analysis
that the absolute value of these solutions have maxima in the
vicinity of classical loci (\ref{eq15}) on the ($u, v$)
configuration space which can exhibit a signature transition
\cite{DOT}. This correspondence between classical loci and quantum
cosmological prediction is considerably interesting. In the
previous section, we have shown that there are duality type
transformations on the parameters of a given scalar field
potential, giving rise to a dual potential, such that the
solutions of the field equations on the ($R,\phi $) configuration
space transform to the dual solutions ($\tilde R,\tilde \phi $),
while the solutions on the ($u, v$) configuration space are {\it
self dual}. Applying the quantum cosmology discussed above to this
picture turns out that for any pair of ($m_1,m_2$) defining a
distinct quantum cosmology in terms of the variables ($u, v$), we
may correspond dual classical solutions ($R, \phi $) and ($\tilde
R,\tilde \phi $), admitting signature transition from Euclidean to
Lorentzian spacetime. This is because, the two sets of dual
classical solutions ($R, \phi $) and ($\tilde R,\tilde \phi $)
have the same quantum cosmology concentrated on both of them over
the configuration space ($u, v$).

\section{Distributional classical solutions}

Now, we consider a manifold with a distribution of the dual
potentials $U$ and $\tilde{U}$. Suppose we have a distribution
\cite {DMT}
\begin{equation}
{\bf U(\phi )}=\Theta ^{+}U^{+}(\phi )+\Theta ^{-}U^{-}(\phi ),
\label{U}
\end{equation}
where $U^{+}\equiv U,U^{-}\equiv \tilde{U}$ and $\Theta
^{+},\Theta ^{-}$ are
Heaviside distributions with support in regions $\beta >0$ (Lorentzian), and $%
\beta <0$ (Euclidean) respectively, such that
\begin{equation}
d\Theta ^{\pm }=\pm \delta,
\end{equation}
where $\delta $ is the hypersurface Dirac distribution with
support on hypersurface $\beta =0$. At the transition hypersurface
$\beta =0$ we assume that the potential is regularly discontinuous
\begin{equation}
[U(\phi )]_{\beta =0}=U^{+}(\phi )\mid _{\beta =0}-U^{-}(\phi
)\mid _{\beta =0},
\end{equation}
where $U^{+}(\phi )\mid _{\beta =0}=lim_{\beta \rightarrow
0^{+}}U(\phi )$ and $U^{-}(\phi )\mid _{\beta =0}=lim_{\beta
\rightarrow 0^{-}}U(\phi )$. This assumption is valid if $\lambda
\neq 0$, otherwise we will have a divergent $\tilde{U}$. In
general, such a manifold is not expected to yield continuous
solutions for the scale factor and the scalar field passing
through the transition hypersurface $\beta =0$, and we obtain
distributional solutions\footnote{Note that, $\dot{R},
\dot{\tilde{R}}, \dot{\phi}$ and $\dot{\tilde{\phi}}$ vanish
approaching the hypersurface $\beta=0$ due to
$\dot{\alpha}(0)=0$.}
\begin{equation}
\Re (\beta )=\Theta ^{+}R^{+}(\beta )+\Theta ^{-}R^{-}(\beta ),
\label{*}
\end{equation}
\begin{equation}
\Phi (\beta )=\Theta ^{+}\phi ^{+}(\beta )+\Theta ^{-}\phi
^{-}(\beta ), \label{**}
\end{equation}
with regularly discontinuous scale factor and scalar field
\begin{equation}
\begin{array}{ll}
\lbrack R(\beta )]_{\beta =0}\neq 0 & ,\hspace{5mm}[\phi (\beta
)]_{\beta =0}\neq 0.
\end{array}
\end{equation}
But, considering Eq.(\ref{eq15}) and self-duality of the parameter
$r=\sqrt{\frac{\lambda _{+}}{\lambda _{-}}}$ the solutions on the
($u, v$) configuration space are smooth and continuous passing
through the hypersurface $\beta=0$, namely
\begin{equation}
[u(\beta )]_{\beta =0}=[v(\beta )]_{\beta =0}=0. \label{+}
\end{equation}
The relevance of this classical model manifests when quantum
cosmology comes in to play the role. We have already seen that
given two disjoint
Lorentzian and Euclidean regions with potentials $U$ and $%
\tilde{U}$ defined over them, respectively, there are classical
solutions ($R(\beta ),\phi (\beta )$) on the former and ($\tilde
R(\beta ),\tilde \phi (\beta )$) on the later one, both of them
corresponding to the same classical cosmology in ($u, v$) space.
On the other hand, as in \cite{DOT} in which a {\it good}
correspondence was shown between the classical loci in $(u, v)$
space and the solutions of the Wheeler-DeWitt equation, one
obtains the same correspondence here. However, unlike \cite{DOT}
where there is a {\it one to one} relation between $(R, \phi)$ and
$(u, v)$, we have here a {\it two to one} relation between $\{(R,
\phi), (\tilde{R}, \tilde{\phi})\}$ and $(u, v)$ classical
cosmologies. Therefore, the solutions of Wheeler-DeWitt equation
in $(u,, v)$ mini-superspace will support not only $(R, \phi)$
cosmology but also its dual, namely $(\tilde{R}, \tilde{\phi})$.
Therefore, following the idea in regarding the Wheeler-Dewitt
solutions more primitive than the classical solutions of Einstein equations \footnote{%
A similar idea was followed by Hawking and Page, investigating the
wormholes \cite{HP}}, we may suggest that to the extent, we are
concerned with quantum cosmology considered here, the
distributions (\ref{*}), (\ref{**}) as combinations of dual
classical cosmologies are also expected to be the allowed
classical predictions . Hence, the strict continuity condition of
the classical fields $R(\beta )$ and $\phi (\beta )$ in passing
through the transition hypersurface $\beta =0$ may be relaxed if
one takes quantum mechanics into account and concentrate on the
{\it good} correspondence between quantum cosmology and dual
classical solutions in this model. In this regard, the jumps of
$R(\beta )$ and $\phi (\beta )$ are {\it quantum cosmologically
allowed effects} which should be interpreted in an appropriate
way.

To the authors knowledge, this is a novel idea in the cosmological
context which may deserve further scrutiny. In particular, it is
of value to know whether this kind of relation between classical
distinguishable and quantum indistinguishable states exists in
non-cosmological contexts. To this end, one may hope to seek such
a relation in any model in which there are some duality
transformations that affect the classical state but leave the
quantum state unchanged.

\section*{Discussion}

Following the model adopted by Dereli and Tucker, we have shown
that in a FRW type manifold endowed completely by $U$ or
$\tilde{U}$, we have signature changing classical solutions $(R,
\phi)$ or $(\tilde{R}, \tilde{\phi})$, corresponding to a same
quantum cosmology. However, the FRW type manifold which we have
assumed here is endowed by a distributional potential (\ref{U})
and correspondingly has distributional solutions with regular
jumps at the hypersurface of signature change. These
distributional solutions are also capable of being the classical
predictions of the quantum cosmology under consideration. One
should then interpret these jumps of classical solutions as a
quantum cosmologically allowed effect. By this effect, we do not
mean at all a quantum transition in the signature of the metric.
Wheeler-DeWitt equation is time ($\beta$ ) independent so is
signature independent, as well. Therefore, both classical limits
in Euclidean and Lorentzian regions are described by a self dual
quantum state $|\psi>_E=|\psi>_L$, and so we have a trivial
quantum transition, namely $_L<\psi|\psi>_E=1$. Of course this is
not surprising because the model is so chosen to represent this
feature. The main point of this model is not to concentrate on the
above ( trivial ) quantum transition from Euclidean to Lorentzian
state, rather the interesting idea is that this self dual quantum
state allows for a reasonable discontinuous jump from one
classical limit in the Euclidean region to the dual one in the
Lorentzian region.

If we regard the configuration space $(u, v)$ as a more primitive
concept than spacetime, we must look for a correspondence between
the properties of the quantum states $\Psi^{(m_1, m_2)}(u, v)$ and
an evolving classical cosmology admitting signature transition. As
is discussed in \cite{DOT}, a change of coordinate $\beta
\rightarrow \beta^\prime=F(\beta)$ for the spacetime patch induces
a change of parametrization for the classical loci (\ref{eq15})
and this corresponds to an alternative choice of the classical
time. To the extent that the classical loci are delineated by a
particular state $\Psi^{(m_1, m_2)}(u, v)$ we may assert that the
family of classical times, regarded as alternative parametrization
of such loci, arise dynamically from this quantum state. From this
point of view, we expect that classical signature change (
$\beta<0 \rightarrow \beta>0$ ) is already included in the
information within $\Psi^{(m_1, m_2)}(u, v)$ \cite{DT}.

According to the scenario for {\it creation of the universe from
nothing}, we know that the universe with a finite size of Planck
length $l_P$ is emerged suddenly, through a quantum tunnelling
effect, from a universe with zero size (nothing) \cite{QT}. This
sudden creation of the {\it finite size} from {\it zero size}
universe is typically an example of jump in the classical
solutions of the Einstein equations, namely from $R=0$ to $R\sim
l_P$, {\it which is allowed by quantum tunnelling}. In the same
way, one may interpret the jumps in the classical solutions
(\ref{*}) and (\ref{**}) as sudden birth of a rather large size
Lorentzian universe from a small size Euclidean region, {\it which
is allowed by quantum cosmology under consideration}.

To this end, considering the transformations (\ref{eq18}) together
with $\xi=S \alpha$ and the inverse transformations
\begin{eqnarray}
R &=& (X^2-Y^2)^{1/3}, \nonumber\\
\label{1516}
 \phi &=& \frac{1}{\alpha}\tanh^{-1}\left(\frac{Y}{X}
\right), \nonumber
\end{eqnarray}
and
\begin{eqnarray}
\tilde{R} &=& (\tilde{X}^2-\tilde{Y}^2)^{1/3}, \nonumber\\
\label{1516'}
 \tilde{\phi} &=&
\frac{1}{\alpha}\tanh^{-1}\left(\frac{\tilde{Y}}{\tilde{X}}
\right), \nonumber
\end{eqnarray}
one finds that the set of parameters
$$\lambda \simeq 0,
\hspace{20mm}m^2 \ll 1, \hspace{20mm} b \simeq 0,
$$
leads to an infinitesimal scale factor in the Euclidean domain and
a finite scale factor in the Lorentzian one \cite{FD}. In the same
way, these parameters map a rather large $|\tilde{\phi}|$ in the
Euclidean domain to the small $|\phi|$ in the Lorentzian
domain\footnote{Considering $\xi=S \alpha$ and (\ref{12}), one may
find that $|X|\gg |\tilde{X}|$ and $Y=\tilde{Y}$, hence we obtain
$|\phi| \ll |\tilde{\phi}|$.}. Exactly, like the quantum
tunnelling scenario \cite{QT} in which the Lorentzian universe
pops out from zero size ( nothing ) to the Planck size, through a
Euclidean {\it instanton} solution, the universe in this simple
model is born at zero size in the Euclidean domain ( instanton
solution ) and evolves to a very small size ( presumably the
Planck size ), approaching the end of Euclidean domain at $\beta
\rightarrow 0^-$. Up to this point, the scenario is in complete
similarity with quantum tunnelling one, because the Euclidean
solutions as instantons ( or real tunnelling solutions ) links the
zero size to the Planck size universe. However, this Planck size
new born universe at the end of Euclidean domain jumps to a finite
size at the beginning of the Lorentzian domain at $\beta
\rightarrow 0^+$. This is the main feature of the model in that
the planck size universe experience a considerable expansion (jump
) in size in a very short period of time. This is reminiscent of
the inflationary scenario where the universe (after quantum
tunnelling) is exponentially expanded from the Planck size to a
finite size in a tiny fraction of a second. Therefore, if we
compare these models and consider the Euclidean {\it instanton}
solution in the present model as the classical description of
quantum tunnelling, then we may interpret the immediate jump in
the scale factor as an inflation-like behavior. It is worth
noticing that this sudden expansion coincides with the signature
transition in the very short interval $\Delta \beta=[\beta
\rightarrow 0^+ - \beta \rightarrow 0^-]$ and ends up at the
beginning of the Lorentzian domain after which the standard model
is applied.

The extent of jumps in the scale factor and the scalar field
depends on the extent of jumps from the parameters
$\{\tilde{\lambda}, \tilde{m}^2, \tilde{b} \}$ to their dual
values $\{ \lambda, m^2, b \}$. On the other hand, due to the
duality relations between the two sets of parameters the extent of
jumps in these parameters depend on the initial values
$\{\tilde{\lambda}, \tilde{m}^2, \tilde{b}\}$. Therefore, the
extent of jump in the scale factor or the scalar field depends on
$\{\tilde{\lambda}, \tilde{m}^2, \tilde{b}\}$. In this regard, the
large values for these parameters, leading to a small effective
cosmological constant $\tilde{\Lambda}_{eff} ( ={\Lambda}_{eff} )$
\cite{FD}, can arrange for a big jump in the scale factor at the
beginning of the Lorentzian region. An effective small
cosmological constant is then playing an important role in the
large expansion of the scale factor at early universe. It is
interesting that, unlike some standard inflationary models with
large cosmological constant, this model predicts an inflation-like
behavior for the scale factor $R$ with a small effective
cosmological constant.

The jump in the scalar field also deserves further scrutiny. The
Euclidean set of large parameters $\{ \tilde{\lambda},
\tilde{m}^2, \tilde{b} \}$ give rise to a large potential
$\tilde{U}(\tilde{\phi})$ in comparison to $U(\phi)$ in terms of
the Lorentzian set of small parameters $\{ \lambda, m^2, b \}$.
Therefore, the jump in the scalar field at the signature
transition hypersurface $\beta =0$ means a jump from the huge
potential $\tilde{U}(\tilde{\phi})$ to the small one $U(\phi)$.
The latter jump may be regarded as a phase transition which {\it
coincides} with the signature transition. The huge energy which is
released during this phase transition may be converted into the
creation of particles in the Lorentzian region to start the hot
big bang. The small parameters $\{ \lambda, m^2, b \}$ in the
Lorentzian region results in a flat potential. On the other hand,
the minimum of the potential $U(\phi)$ occurs at the negative
value $\phi=-\frac{1}{2\alpha}\tanh^{-1}(\frac{2b}{m^2})\ll 0$.
According to the signature dynamics however, the negative $\phi\ll
0$ is realized in the Lorentzian region at $\beta \gg 0$, which
means that the minimum of the potential is realized at $\beta \gg
0$. Therefore, after phase transition at $\beta =0$ the scalar
field in the Lorentzian side rolls down to its minimum at $\beta
\gg 0$ very slowly due to the flat potential and releases its
remnant energy into the universe at a small rate.

In conclusion, it is worth noting that the correspondence between
the quantum states and the classical loci is independent of the
following scale transformations
$$
\lambda \rightarrow \xi \lambda \:\:\:,\:\:\: m^2 \rightarrow \xi
m^2 \:\:\:,\:\:\: b \rightarrow \xi b,
$$
on the parameter space, leading to $\lambda_{\pm} \rightarrow \xi
\lambda_{\pm}$ and $r \rightarrow r$. These leave the classical
loci Eq.(\ref{eq15}) and the quantization condition (\ref{123})
unchanged and do not affect the general patterns of the quantum
amplitude $|\Psi^{(m_1, m_2)}(u,v)|^2$ \cite{DOT}. This indicates
that a class of universes may be defined by the equivalency
relation
$$
U(\phi) \rightarrow \xi U(\phi),
$$
in which a good correspondence exists between the quantum states
and the classical loci, for each universe. This, on the other
hand, shows that such a correspondence exists for an equivalency
class of universes with the effective cosmological constants
related to each other by
$$
\Lambda_{eff} \rightarrow \xi \Lambda_{eff}.
$$
Note that, although the above scale transformations do not affect
the classical loci, but change the domain of the Euclidean region
\cite{DOT}, in a parametric way, as
$$
-(3\pi/2r\omega_2)^{2/3}\leq\beta<0 \longrightarrow
-(3\pi/2r\omega_2)^{2/3} \xi^{-1/3}\leq\beta<0
$$
while the domain of the Lorentzian region $0<\beta\leq\infty$
remains unchanged. Therefore, each universe in the above
equivalency class has its own Euclidean region whose domain is
different from other one.

\section*{Acknowledgment}

We would like to thank the anonymous referee whose useful comments
improved the paper. This work has been financially supported by
the Research department of Azarbaijan University of Tarbiat
Moallem, Tabriz, Iran.

\end{document}